\documentclass[]{aastex631}

\usepackage{bm}
\usepackage{CJK}
\usepackage{soul}

\shorttitle{Cosmic tidal reconstruction in redshift space}
\shortauthors{Zang et al.}
\graphicspath{{./}{figures/}}

\begin{document}

\title{Cosmic tidal reconstruction in redshift space}


\author[0000-0003-2299-6235]{Shi-Hui Zang \begin{CJK*}{UTF8}{gbsn}(臧诗慧)\end{CJK*}}
\affiliation{School of Aerospace Engineering, Tsinghua University, 30 Shuangqing Road, Beijing 100084, China}

\author[0000-0002-8202-8642]{Hong-Ming Zhu \begin{CJK*}{UTF8}{gbsn}(朱弘明)\end{CJK*}}
\affiliation{Canadian Institute for Theoretical Astrophysics, 60 St. George Street, Toronto, Ontario M5S 3H8, Canada}

\author{Marcel Schmittfull}
\affiliation{PDT, 1745 Broadway, New York, NY 10019, USA}
\affiliation{School of Natural Sciences, Institute for Advanced Study, 1 Einstein Drive, Princeton, NJ 08540, USA}

\author[0000-0003-2155-9578]{Ue-Li Pen \begin{CJK*}{UTF8}{bsmi}(彭威禮)\end{CJK*}}
\affiliation{Institute of Astronomy and Astrophysics, Academia Sinica, Astronomy-Mathematics Building, No. 1, Sec. 4, Roosevelt Road, Taipei 10617, Taiwan}
\affiliation{Canadian Institute for Theoretical Astrophysics, 60 St. George Street, Toronto, Ontario M5S 3H8, Canada}
\affiliation{Dunlap Institute for Astronomy and Astrophysics, University of Toronto, 50 St. George Street, Toronto, Ontario M5S 3H4, Canada}
\affiliation{Canadian Institute for Advanced Research, 661 University Avenue, Toronto, Ontario M5G 1M1, Canada}
\affiliation{Perimeter Institute for Theoretical Physics, 31 Caroline Street North, Waterloo, Ontario N2L 2Y5, Canada}

\begin{abstract}

The gravitational coupling between large- and small-scale density perturbations leads to anisotropic distortions to the local small-scale matter fluctuations.
Such local anisotropic distortions can be used to reconstruct the large-scale matter distribution, known as tidal reconstruction.
In this paper, we apply the tidal reconstruction methods to simulated galaxies in redshift space.
We find that redshift-space distortions lead to anisotropic reconstruction results.
While the reconstructed radial modes are more noisy mainly due to the small-scale velocity dispersion, the transverse modes are still reconstructed with high fidelity, well correlated with the original large-scale density modes.
The bias of the reconstructed field at large scales shows a simple angular dependence which can be described by a form similar to the linear redshift-space distortion.
The noise power spectrum is nearly isotropic and scale-independent on large scales.
This makes the reconstructed tides field an ideal tracer for cosmic variance cancellation and multi-tracer analysis and has profound implications for future 21~cm intensity mapping surveys.
\end{abstract}


\section{Introduction}
\label{sec:introduction}

Galaxy peculiar velocities contribute to a galaxy's observed redshift via the Doppler effect.
This leads to characteristic anisotropies in the observed galaxy clustering pattern, known as redshift-space distortions (RSDs) \citep[][]{1972MNRAS.156P...1J,1977ApJ...212L...3S,1980lssu.book.....P,1987MNRAS.227....1K,1994MNRAS.267.1020P,1996MNRAS.282..877B}.
By measuring the velocity-induced statistical effect on the galaxy power spectrum, recent galaxy surveys have measured the structure growth rate accurately, providing constraints on both dark energy properties and modifications to gravity theory \citep[e.g. SDSS BOSS/eBOSS][]{2017MNRAS.470.2617A,2021PhRvD.103h3533A}.
The forthcoming generation of wide-field galaxy redshift surveys generally probe larger volumes and higher galaxy densities, thus allowing for higher signal to noise ratio measurements and better insights into our Universe, e.g. DESI \citealt{2016arXiv161100036D}, PFS \citep[][]{2014PASJ...66R...1T}, Euclid \citep[][]{2018LRR....21....2A,2020A&A...642A.191E}, SPHEREx \citep[][]{2014arXiv1412.4872D}, LSST \citep[][]{2009arXiv0912.0201L}, MegaMapper \citep[][]{2019BAAS...51g.229S,2019BAAS...51c..72F}.

An accurate theoretical description for galaxy clustering in redshift space is a key for the success of the future spectroscopic surveys \citep[see e.g.][]{2020PhRvD.102l3541N}.
However, the modeling of the galaxy power spectrum is limited to $k\sim0.14\ h\mathrm{Mpc}^{-1}$, mainly due to the finger of God effect on small scales \citep[see e.g.][]{2021arXiv211000006I,2021arXiv211000016D}.
Simulation-based methods can in principle capture the non-linear RSD effects, but recently \citet{2021arXiv211006969K} find little improvement in cosmological parameters beyond $k\sim0.2\ h\mathrm{Mpc}^{-1}$, due to the degeneracy between cosmological parameters and nuisance parameters in the analysis.
The stage-III spectroscopic surveys can already map the galaxy distributions to $k\sim0.2\ h\mathrm{Mpc}^{-1}$, and will be substantial to $k\sim1\ h\mathrm{Mpc}^{-1}$ by future spectroscopic surveys.
Thus, it is necessary to develop new methods to exploit small-scale information in redshift space.

The gravitational coupling between density perturbations leads to striking non-Gaussian features in the large-scale structure.
The small-scale filamentary structures arise from gravitational tidal interactions.
The gravitational nonlinearity has traditionally led to a reduction in cosmological information \citep[e.g.][]{1999MNRAS.308.1179M,2005MNRAS.360L..82R}.
It has been realized that such tidal non-Gaussianity can be exploited to improve the measurement of large-scale structures \citep[][]{2012arXiv1202.5804P}.
The local anisotropic distortions can be used to reconstruct the large-scale tidal shear and gravitational potential \citep[][]{2012arXiv1202.5804P,2016PhRvD..93j3504Z,2022ApJ...929....5Z}.
\citet{2012arXiv1202.5804P} presented the first tidal reconstruction method, which uses two transverse shear fields in analogy with weak lensing \citep[][]{1993ApJ...404..441K}.
This method has been further explored by \citet{2016PhRvD..93j3504Z} and found to be noisier for radial modes along the line of sight, i.e. an anisotropic reconstruction noise.
This is because these modes are inferred indirectly from the variations of the two transverse shear fields along the line of sight.
A new algorithm which exploits all five shear terms in three-dimensional space has been proposed recently, which reconstructs the radial modes directly from another three shear fields \citep{2022ApJ...929....5Z}.
The new method has a lower and isotropic reconstruction noise, compared to the previous method using two shear fields.
Similar algorithms have also been investigated by other groups, following the nonlinear coupling in standard perturbation theory \citep[see][for more details]{2018JCAP...07..046F,2020PhRvD.101h3510L,2020arXiv200700226L,2021PhRvD.104l3520D}. 

The reconstructed field from the tidal effects provides an independent tracer of the large-scale structure.
By comparing this to redshift space galaxy field, one can measure the velocity growth factor on large scales without cosmic variance, analogous to \citet{2009JCAP...10..007M}.
This enables precision measurements of local-type primordial non-Gaussianity using an effective multi-tracer approach \citep[see][for more discussions]{2021PhRvD.104l3520D}, where the improvements arise from the cosmic variance cancellation \citep[][]{2009MT}.

In 21~cm cosmology, the radial modes with small wave numbers are lost due to the Galactic foreground contamination.
However, these modes can be reconstructed with tidal reconstruction
\citep[][]{2018PhRvD..98d3511Z,2018JCAP...07..046F,2019PhRvD.100b3517L,2019MNRAS.486.3864K}.
This is essential for CMB and other correlations such as weak lensing, kinematic Sunyaev-Zel'dovich effect, photometric galaxies, etc \citep[see e.g.][]{2018PhRvD..98d3511Z,2019PhRvD.100b3517L,2021arXiv211205034G}.
The recovery of 21~cm radial modes opens up a new set of possibilities and has profound implications for 21~cm cosmology.
This problem has also been explored by \citet{2019JCAP...11..023M,2021JCAP...10..056M} and \citet{2021MNRAS.504.4716G} using a forward modeling approach and a machine learning-based method, which could be more optimal with a higher computational cost.

These applications rely on a successful implementation of tidal reconstruction in redshift space, while previous studies have focused on tidal reconstruction in real space.
Measurements of density fields from galaxies are exclusively made in redshift space. In principle, the RSD effect can be included in the nonlinear coupling in standard perturbation theory, \citep[e.g. following][]{2018JCAP...07..046F,2020PhRvD.101h3510L,2021PhRvD.104l3520D,2020arXiv200700226L}, which should be valid in the mildly nonlinear regime.
However, galaxies are subject to nonlinear dynamics.
While the leading order effect is well described by perturbation theory on large scales, the nonlinear effects on small scales, i.e. fingers of God, are difficult to model, limiting an analytical study of redshift space tidal reconstruction.

In this work, we present a detailed study of tidal reconstruction in redshift space.
We apply the tidal reconstruction methods to mock galaxies from $N$-body simulations.
We find that the reconstruction results are anisotropic due to the RSD effect.
While the radial modes are more noisy due to the nonlinear velocity dispersion, the transverse modes can be reconstructed with high fidelity, well correlated with the large-scale matter density field.
The large-scale bias of the reconstructed field can be described by a simple two-parameter model with a distinct angular dependence to the linear RSD effect and the noise power spectrum is nearly isotropic and scale-independent on large scales, which can be relatively straightly fitted in the cosmological parameter inference.
This enables tidal reconstruction a promising method for the multi-tracer analysis and 21~cm intensity mapping surveys.
    
This paper is organized as follows. 
In Section~\ref{sec:tidalreconstruction}, we introduce the tidal reconstruction methods. 
In Section~\ref{sec:method}, we describe the numerical simulations and the numerical implementation of tidal reconstruction.
In Section~\ref{sec:result}, we present the numerical results of reconstruction.
We discuss and conclude in Section~\ref{sec:discussion}.

\section{METHODOLOGY}
\label{sec:tidalreconstruction}

The gravitational coupling between large- and small-scale perturbations leads to anisotropic distortions in the locally measured correlation function \citep[][]{2012arXiv1202.5804P,2010PhRvL.105p1302M,2012PhRvL.108y1301J}.
Such local anisotropic tidal distortions can be used to reconstruct the large-scale matter distribution \citep[][]{2012arXiv1202.5804P,2016PhRvD..93j3504Z,2022ApJ...929....5Z}.
In this section, we present the tidal reconstruction algorithm and discuss its redshift space application.

We consider the gravitational interaction between a long wavelength perturbation and small-scale density fluctuations in the squeezed limit, i.e., the wavelength of the small-scale density fluctuations is sufficiently smaller than that of the large-scale density field.
The leading order observable is then described by the large-scale tidal field,
\begin{equation}
    t_{ij}=\Phi_{L,ij},
\end{equation}
where $\Phi_L$ is the large-scale gravitational potential.
The $3\times3$ symmetric tensor field $t_{ij}$ can be decomposed as 
\begin{equation}
\label{eq:tij}
    t_{ij} = \left(
    \begin{array}{ccc}
    \epsilon_0 + \epsilon_1 - \epsilon_z & \epsilon_2 & \epsilon_x \\
    \epsilon_2 & \epsilon_0 -\epsilon_1 - \epsilon_z & \epsilon_y \\
    \epsilon_x & \epsilon_y &  \epsilon_0 + 2\epsilon_z
    \end{array}
    \right),
\end{equation}
where $\epsilon_{0}=(\Phi_{L,11}+\Phi_{L,22}+\Phi_{L,33})/3$, $\epsilon_1 = (\Phi_{L, 11} - \Phi_{L, 22} )/2, \epsilon_2 = \Phi_{L, 12}, \epsilon_x = \Phi_{L, 13}, \epsilon_y = \Phi_{L, 23}$ and $\epsilon_z = (2\Phi_{L, 33} - \Phi_{L, 11} - \Phi_{L,22})/6$.
The trace part of the tidal field corresponds to the local mean density, while other components describe the tidal shear terms.
The gravity shear forces lead to anisotropic distortions in the locally measured power spectrum \citep[see e.g.][for more details]{2014PhRvD..89h3507S}.
Since the large-scale tidal field is coherent on small scales, the tidal coupling results in a systematic change of the small-scale power.
When enough small-scale modes are measured, the tidal shear terms can be reconstructed with high fidelity \citep[][]{2012arXiv1202.5804P}.

The large-scale tidal shear fields can be estimated with the quadratic 
estimators, which are outer products of the filtered density fields \citep{2008MNRAS.388.1819L, 2010PhRvD..81l3015L, 2012PhRvD..85d3016B},
\begin{eqnarray}
\label{eq:shearestimation}
    \hat{\epsilon}_1(\bm{x}) & = & [\delta^{w_1}(\bm{x})\delta^{w_1}(\bm{x}) - \delta^{w_2}(\bm{x})\delta^{w_2}(\bm{x})]/2, \nonumber \\
    \hat{\epsilon}_2(\bm{x}) & = & \delta^{w_1}(\bm{x})\delta^{w_2}(\bm{x}), \nonumber \\
    \hat{\epsilon}_x(\bm{x}) & = & \delta^{w_1}(\bm{x})\delta^{w_3}(\bm{x}), \nonumber \\
    \hat{\epsilon}_y(\bm{x}) & = & \delta^{w_2}(\bm{x})\delta^{w_3}(\bm{x}), \nonumber \\
    \hat{\epsilon}_z(\bm{x}) & = & [2\delta^{w_3}(\bm{x})\delta^{w_3}(\bm{x}) - \delta^{w_1}(\bm{x})\delta^{w_1}(\bm{x}) \nonumber \\
    &&- \delta^{w_2}(\bm{x})\delta^{w_2}(\bm{x})]/6,
\end{eqnarray}
where
\begin{equation}
    \label{eq:filterestimation}
    \delta^{w_j}(k)=ik_j W_R(k)\delta(\bm{k}),
\end{equation}
is the filtered gradient density field and $W_R(k) = \exp(-k^2R^2/2)$ is the Gaussian window with smoothing scale $R$ \citep[][]{2022ApJ...929....5Z}.

In principle, the filter here should be anisotropic in redshift space. 
The mapping from real to redshift space brings anisotropies in the observed galaxy distribution along the line of sight, including the Kaiser effect \citep[][]{1987MNRAS.227....1K} and fingers of God \citep[][]{1972MNRAS.156P...1J}.
The radial modes at small scales are usually noisier due to the fingers of God damping \citep[][]{2021JCAP...05..059S} and anisotropic smoothing can account for this and improve the performance
\citep[][]{2016MNRAS.457.2068C,2018MNRAS.478.1866H}.
However, it is difficult to quantify the impact of RSD on reconstruction when using an anisotropic smoothing window as we are observing the combined effects of RSD and anisotropic filtering.

The real to redshift space mapping also causes additional coupling of small-scale densities to the large-scale density field along the line of the sight and this can be computed using perturbation theory \citep[][]{2017PhRvD..95h3522A,2017JCAP...06..053B,2018JCAP...02..022L,2018PhRvD..97f3527A,2018JCAP...07..049C,2019PhRvD.100j3515A}.
The estimators can be constructed using the standard perturbation theory in redshift space to account for the anisotropic coupling due to the RSD effects, which potentially enables an unbiased estimate of the real space large-scale matter field \citep[e.g. following methods in][]{2018JCAP...07..046F,2020PhRvD.101h3510L,2020arXiv200700226L,2021PhRvD.104l3520D}.
However, this limits the number of small-scale modes that can be included in reconstruction as perturbation theory is only valid in the mildly nonlinear regime, which degrades the reconstruction significantly \citep[see][for more discussions]{2022ApJ...929....5Z}.

With the estimated tidal shear fields, we construct estimators for the large-scale density field.
In general, any combination of shear fields can provide an estimate of the large-scale density field \citep[][]{2022ApJ...929....5Z}.
Here, we consider two tidal reconstruction algorithms.
One uses two transverse shear fields, $\epsilon_1$ and $\epsilon_2$, which are less affected by errors in redshift estimation \citep{2012arXiv1202.5804P, 2016PhRvD..93j3504Z,2019MNRAS.486.3864K}.
Another algorithm exploits all five shear terms and thus has a lower reconstruction noise \citep{2022ApJ...929....5Z}. 
The details of the two methods are outlined below.
\begin{itemize}    
\item {\it Transverse shear reconstruction}:
In \citet{2012arXiv1202.5804P} and \citet{2016PhRvD..93j3504Z}, we uses two purely transverse shear field $\epsilon_1$ and $\epsilon_2$ in analogy with the weak-lensing mass reconstruction \citep[][]{1993ApJ...404..441K}.
The large-scale density field is given by
\begin{equation}
\label{eq:2shear}
\epsilon_0(\bm{k}) = \frac{2k^2}{3(k_1^2 + k_2^2)^2} \left[ (k_1^2 - k_2^2)\epsilon_1(\bm{k}) + 2k_1 k_2 \epsilon_2(\bm{k}) \right],
\end{equation}
where $\epsilon_0=\nabla^2\Phi_L/3$, which differs from the large-scale density $\delta_L$ by a constant proportional factor.
This original proposal can avoid the impact of RSD on reconstruction since the transverse tidal shears in the tangential plane are less sensitive to the RSD effect along the line of sight.
The RSD effect should be a second order effect for reconstruction.
In this paper, we explore the redshift space performance of this method in detail.

\item {\it Full shear reconstruction}: This method is proposed by \citet{2022ApJ...929....5Z}, where we exploit all five shear terms in reconstruction. 
The reconstructed field is given by
\begin{eqnarray}
\label{eq:5shear}
\epsilon_0(\bm{k}) =  \frac{1}{2k^2} & \left[ (k_1^2 - k_2^2)\epsilon_1(\bm{k}) + 2k_1k_2\epsilon_2(\bm{k}) + 2k_1k_3\epsilon_x(\bm{k}) \right .\\ 
    &  \left . + 2k_2k_3\epsilon_y(\bm{k}) + (2k_3^2 - k_1^2 - k_2^2)\epsilon_z(\bm{k}) \right]. \nonumber
\end{eqnarray}
The full shear reconstruction uses full shear information in the three-dimensional space and thus has a lower and isotropic reconstruction noise in real space \citep[][]{2022ApJ...929....5Z}.
However, the shear fields $\epsilon_x$, $\epsilon_y$ and $\epsilon_z$ directly probe the inhomogeneous matter distribution in the line of sight direction, which is affected by the mapping from real to redshift space.
Therefore, it is expected that the reconstruction will be highly anisotropic compared with transverse shear reconstruction in redshift space.
We explore the detailed performance in redshift space with simulated mock galaxy fields below.
\end{itemize}

In general, the reconstructed density field can be written as 
\begin{equation}
    \label{eq:model2}
    \delta_r(\bm{k}) = T(\bm{k}) \delta(\bm{k})+ N(\bm{k}),
\end{equation}
where $\delta_r(\bm{k})$ denotes the reconstructed field from transverse or full shear algorithm, $T(\bm{k})$ is the propagator that quantifies the bias to the original real space dark matter density field $\delta(\bm{k})$ and $N(\bm{k})$ is reconstruction noise.
For full shear reconstruction in real space, the reconstruction bias and noise power only depend on the magnitude of the wave vector \citep[][]{2022ApJ...929....5Z}.
However, for tidal reconstruction in redshift space, we expect that both the propagator $T(\bm{k})$ and the noise $N(\bm{k})$ will depend on the magnitude of the wave vector as well as the angle between the wave vector and the radial direction.
We explore the properties of the propagator and noise power using high precision simulations in the following sections.

\section{Numerical setup}
\label{sec:method} 

In this section, we describe the simulations and the numerical implementation of tidal reconstruction.

\subsection{Simulations}
\label{sec:datasample}
To investigate the performance of tidal reconstruction in redshift space, we utilize a set of six independent $N$-Body simulations, run with {\tt MP-Gadget} \citep{feng2018}, evolving $1536^3$ dark matter particles in a periodic box with side length $L=1500\ h^{-1}\mathrm{Mpc}$ to redshift $z=0.6$.
The cosmological parameters are 
$\Omega_{m} = 0.3075$, $\Omega_bh^2=0.0223$, $\Omega_ch^2=0.1188$, $h = 0.6774$, $\sigma_8=0.8159$, and $n_s=0.9667$.
These are the same simulations used in \citet{2021JCAP...05..059S} and \citet{2021JCAP...03..020S}. 

The {\tt Rockstar} \citep{2013ApJ...762..109B} phase space halo finder is used to identify halos and subhalos from snapshots of dark matter particles at redshift $z=0.6$.
We generate the simulated galaxy samples by imposing a soft mass cut to the viral mass to select massive halos and subhalos to represent galaxies, following the procedure of \citet{2020PhRvD.102l3541N}.
There are two parameters, $\log_{10}M_{\mathrm{min}}$ and $\sigma_{\log_{10}M}$, which determine typical minimum mass and the profile of the soft mass cutoff \citep{2020PhRvD.102l3541N}.
By choosing $\log_{10}M_{\mathrm{min}}=11.5\  h^{-1}M_{\odot}$ and $12.97\  h^{-1}M_{\odot}$, we have two galaxy samples with number densities $\bar{n}=3.6\times10^{-3}\ h^3\mathrm{Mpc}^{-3}$ and $4.25\times10^{-4}\ h^3\mathrm{Mpc}^{-3}$, respectively.
The value of $\sigma_{\log_{10}M}$
is 0.35 for both samples.
See \citet{2020PhRvD.102l3541N} and \citet{2021JCAP...05..059S} for more details.
The higher mass sample approximately reproduces the observed properties of BOSS CMASS galaxies \citep[][]{2020PhRvD.102l3541N} and the lower mass mock galaxies that will be observed by DESI have a much higher number density \citep{2021JCAP...05..059S}.
We use these two catalogs to explore the effects of number density on tidal reconstruction. 

We implement the RSD by moving galaxies along the line of sight according to the velocities of center-of-mass given by {\tt Rockstar}.
The redshift space position $\bm{s}$ of a galaxy at true comoving position $\bm{x}$ is given by
\begin{equation}
    \label{eq:rsd}
    \bm{s} = \bm{x} + \frac{\hat{\bm{z}}\cdot\bm{v}(\bm{x}) }{aH}\hat{\bm{z}},
\end{equation}
where $\bm{v}$ is the peculiar velocity, $a$ is the scale factor and $H$ the Hubble parameter.
The logarithmic structure growth rate at $z=0.6$ is $f=0.786$.
In this work, we adopt the plane-parallel or distant-observer approximation, in which the $\hat{\bm{z}}$ direction is taken to be the line of sight.

We use the standard (Cloud-in-Cell) CIC interpolation scheme to paint galaxies and particles to a $1536^3$ regular grid.
The resulting density fields are deconvolved with the CIC window \citep{2005ApJ...620..559J}. 
We also use interlacing to reduce the effect of aliasing caused by the finite sampling \citep{2016MNRAS.460.3624S}.

The redshift space analysis requires a separate treatment of line-of-sight and transverse components of $\bm{k}$.
We compute the two-dimensional power spectrum of density fields, $P(k_\perp,k_\parallel)$, where $k_\perp$ and $k_\parallel$ are the transverse and line-of-sight components of $\bm{k}$.
For better quantitative assessments of the reconstruction performance, the power spectra of density fields are also computed in discrete $k$ and $\mu$ bins, where $k$ is the magnitude of the wave vector and $\mu=k_{\parallel}/k$ is the cosine of the angle between the line-of-sight and the wave vector.
We use five uniform $\mu$ bins, $\mu=0-0.2,0.2-0.4$, etc.
The width of $k$ bins is $\Delta k=3k_f$ for both $P(k_\perp,k_\parallel)$ and $P(k,\mu)$, where
$k_f$ is the fundamental frequency $k_f = 2\pi/L$.
In the following discussions, we use $P_{AB}(\bm{k})\equiv\langle A(\bm{k})B^*(\bm{k})\rangle$ to denote the power spectrum of fields $A(\bm{k})$ and $B(\bm{k})$. 
Note that here we have dropped the Dirac delta function.

\subsection{Reconstruction}
\label{sec:reconstruction}

The tidal reconstruction works as follows.
We first smooth the redshift space galaxy field with the Gaussian window and compute the filtered field using Equation~(\ref{eq:filterestimation}).
The optimal smoothing scales are different for the two galaxy samples.
We test a few different scales and the optimal scales which maximize the correlation are $1.25\ h^{-1}\mathrm{Mpc}$ and $1.5\ h^{-1}\mathrm{Mpc}$ for galaxy samples with $\bar{n}=3.6\times10^{-3}\ h^3\mathrm{Mpc}^{-3}$ and $4.25\times10^{-4}\ h^3\mathrm{Mpc}^{-3}$, respectively.
Then the shear fields are computed following Equation~(\ref{eq:shearestimation}).
Finally, the reconstructed large-scale density is given by Equation~(\ref{eq:2shear}) and Equation~(\ref{eq:5shear}) for transverse and full shear reconstruction.

To better analyse the reconstruction results,
we have written the reconstructed field as
\begin{equation}
    \delta_r(\bm{k}) = T(\bm{k})\delta(\bm{k}) + N(\bm{k}),
\end{equation}
where $T(\bm{k})=P_{\delta_r\delta}(\bm{k})/P_{\delta\delta}(\bm{k})$ is the propagator, the $\delta(\bm{k})$ is dark matter field in real space, and $N(\bm{k})$ is the reconstruction noise.
Note that the power spectrum of dark matter density field $P_{\delta\delta}$ is isotropic in real space due to statistical isotropy.
Therefore, the propagator $T(k_\perp,k_\parallel)$ or $T(k,\mu)$ fully quantifies the angular dependence of the reconstructed field due to the RSD effect.

The power spectrum of the reconstruction error, or noise, which describes the stochasticity for tidal reconstruction, is given by
\begin{equation}
    \label{eq:noisepower}
    P_{\mathrm{err}}(\bm{k}) \equiv \langle|\delta_r(\bm{k}-T(\bm{k})\delta(\bm{k})|^2\rangle = \left( P_{\delta_r\delta_r}(\bm{k}) - \frac{P_{\delta_r\delta}(\bm{k})^2}{P_{\delta\delta}(\bm{k})} \right),
\end{equation}
where we have used $T(\bm{k})=P_{\delta_r\delta}(\bm{k})/P_{\delta\delta}(\bm{k})$.
It is natural to expect that the noise power spectrum would be anisotropic for tidal reconstruction in redshift space.

We also compute the cross correlation coefficient between the reconstructed field and the real space dark matter density field,
\begin{equation}
    \label{eq:crosscorrelation}
    r_{cc}(\bm{k}) = \frac{P_{\delta_r\delta}(\bm{k})}{\sqrt{P_{\delta\delta}(\bm{k})P_{\delta_r\delta_r}(\bm{k})}},
\end{equation}
where $P_{\delta_r\delta}(\bm{k})$ is the cross power spectrum, and $P_{\delta\delta}(\bm{k})$ and $P_{\delta_r\delta_r}(\bm{k})$ are the power spectra of the real space dark matter density $\delta$ and reconstructed field $\delta_r$. 
Higher correlation between the two fields indicates a better reconstruction.

Obviously in a perfect reconstruction we have $r(\bm{k})=1$, and $P_{\mathrm{err}}(\bm{k}) = 0$.
From Equation~(\ref{eq:noisepower}) and Equation~(\ref{eq:crosscorrelation}), it can be derived that the noise power spectrum divided by the total power spectrum of reconstructed field is related to the cross correlation coefficient as
\begin{equation}
\label{eq:1-r2}
    P_{\mathrm{err}}(\bm{k})/P_{\delta_r\delta_r}(\bm{k})=1-r^2_{cc}(\bm{k}).
\end{equation}

For optimal filtering the reconstructed fields, we can compute the transfer function by minimizing the difference between the reconstructed and real space dark matter density fields,
\begin{equation}
    \langle|t(\bm{k})\delta_r(\bm{k})-\delta(\bm{k})|^2\rangle,
\end{equation}
and we have 
\begin{equation}
\label{eq:tf}
    t(\bm{k})=\frac{P_{\delta_r\delta}(\bm{k})}{P_{\delta\delta}(\bm{k})}.
\end{equation}
Note that for tidal reconstruction in redshift space, the transfer function depends on the cosine $\mu$, as the reconstruction noise is highly anisotropic.

The power spectra are measured in $(k_\perp,k_\parallel)$ or $(k,\mu)$ bins for each simulation first and then averaged over the six independent realizations to suppress cosmic variance before computing $T(\bm{k})$, $r_{cc}(\bm{k})$, and $t(\bm{k})$.

\section{RESULTS}
\label{sec:result} 

In this section, we assess the performance of two tidal reconstruction algorithms with simulated galaxy mock catalogs from high precision simulations.
We consider several metrics including the density maps, the cross-correlation coefficient with the real space dark matter density field, the propagator and the noise power spectrum.
We then turn to exploring the angular-dependent reconstruction effects with synthetic redshift space galaxy samples.

\subsection{Full shear reconstruction}

\begin{figure}[ht!] 
    \centering
    \includegraphics[width=\columnwidth]{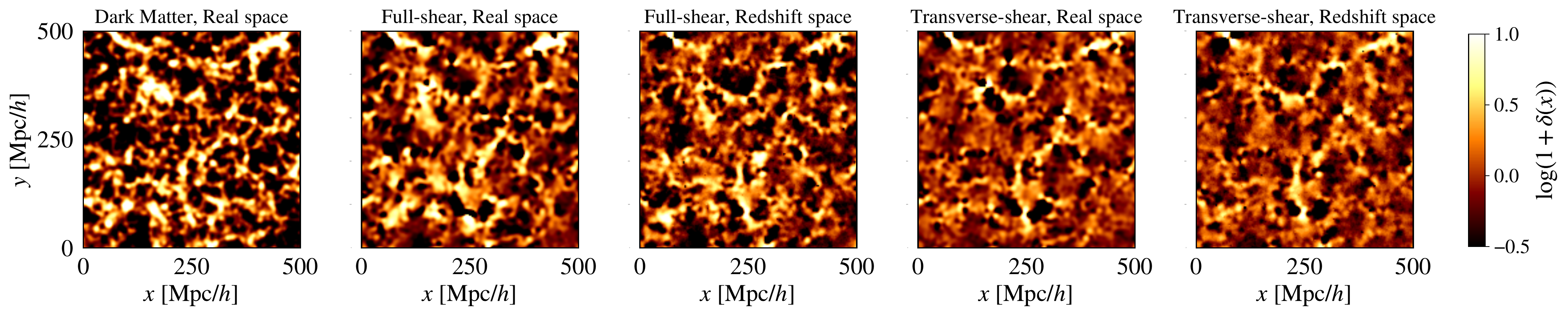}
    \caption{Two-dimensional slices of the density maps in the $x-y$ plane.
    From left to right, the panels show
    the dark matter density field, the full shear tidal reconstructed field in real and redshift spaces and the corresponding reconstructed fields for the transverse method.
    The maps are reconstructed from the lower mass catalog with number density $\bar{n} = 3.60\times 10^{-3}\ h^3\mathrm{Mpc}^{-3}$.
    The reconstructed fields are convolved with the transfer function in Equation~(\ref{eq:tf}) to suppress the anisotropic noises.
    The dark matter density field is smoothed by a Gaussian filter with smoothing scale $R = 4\ h^{-1}\mathrm{Mpc}$. 
    }
    \label{fig:Slice_5T_XY}
\end{figure}

Figure~\ref{fig:Slice_5T_XY} shows the two-dimensional slices of the dark matter density field in one of the simulations, and the two fields reconstructed with the lower mass galaxy density fields in real and redshift space, respectively.
The number density of this catalog is $\bar{n} = 3.60\times 10^{-3}\ h^3\mathrm{Mpc}^{-3}$.
The dark matter density field is smoothed with an $R=4\ h^{-1}\mathrm{Mpc}$ Gaussian.
The reconstructed fields are convolved with the transfer function in Equation~(\ref{eq:tf}), minimizing the difference between the reconstructed field and the real space dark matter density field.
This process effectively corrects the anisotropic bias and suppresses the anisotropic noise to have a better visual comparison.
We see that tidal reconstruction can provide an accurate estimate of the large-scale matter distribution, consistent with findings in \citet{2022ApJ...929....5Z}.
The redshift space reconstruction shows similar performance as the real space result, i.e., the RSD does not impact the reconstruction in the transverse plane much.
This is not surprising as the RSD effect mainly changes the line of sight modes, and does not affect transverse modes with $\mu\simeq0$ a lot.

\begin{figure}[ht!] 
    \centering
    \includegraphics[width=\columnwidth]{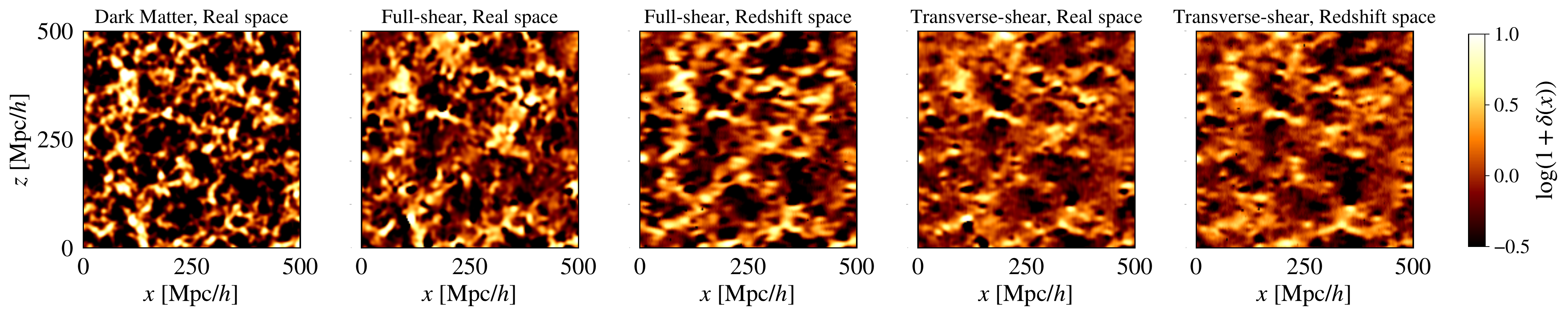}
    \caption{Same as Fig.~\ref{fig:Slice_5T_XY}, but for density slices in the $x-z$ plane.}
    \label{fig:Slice_5T_XZ}
\end{figure}

Figure~\ref{fig:Slice_5T_XZ} compares the full shear reconstructed fields with the real space dark matter density field in the $x-z$ plane.
In contrast to Figure~\ref{fig:Slice_5T_XZ}, these density slices directly probe the performance of tidal reconstruction in the radial direction.
As expected, the reconstructed field shows an anisotropic noise in the $x-z$ plane.
The reconstructed map is noisier than the corresponding real space map.
We expect that the reconstruction degrades mainly due to the small-scale nonlinear RSD effect, i.e., fingers of God, since tidal reconstruction is dominated by the large number of small-scale modes \citep[][]{2022ApJ...929....5Z}.


\begin{figure}[ht!]
    \centering
    \includegraphics[width=0.7\columnwidth]{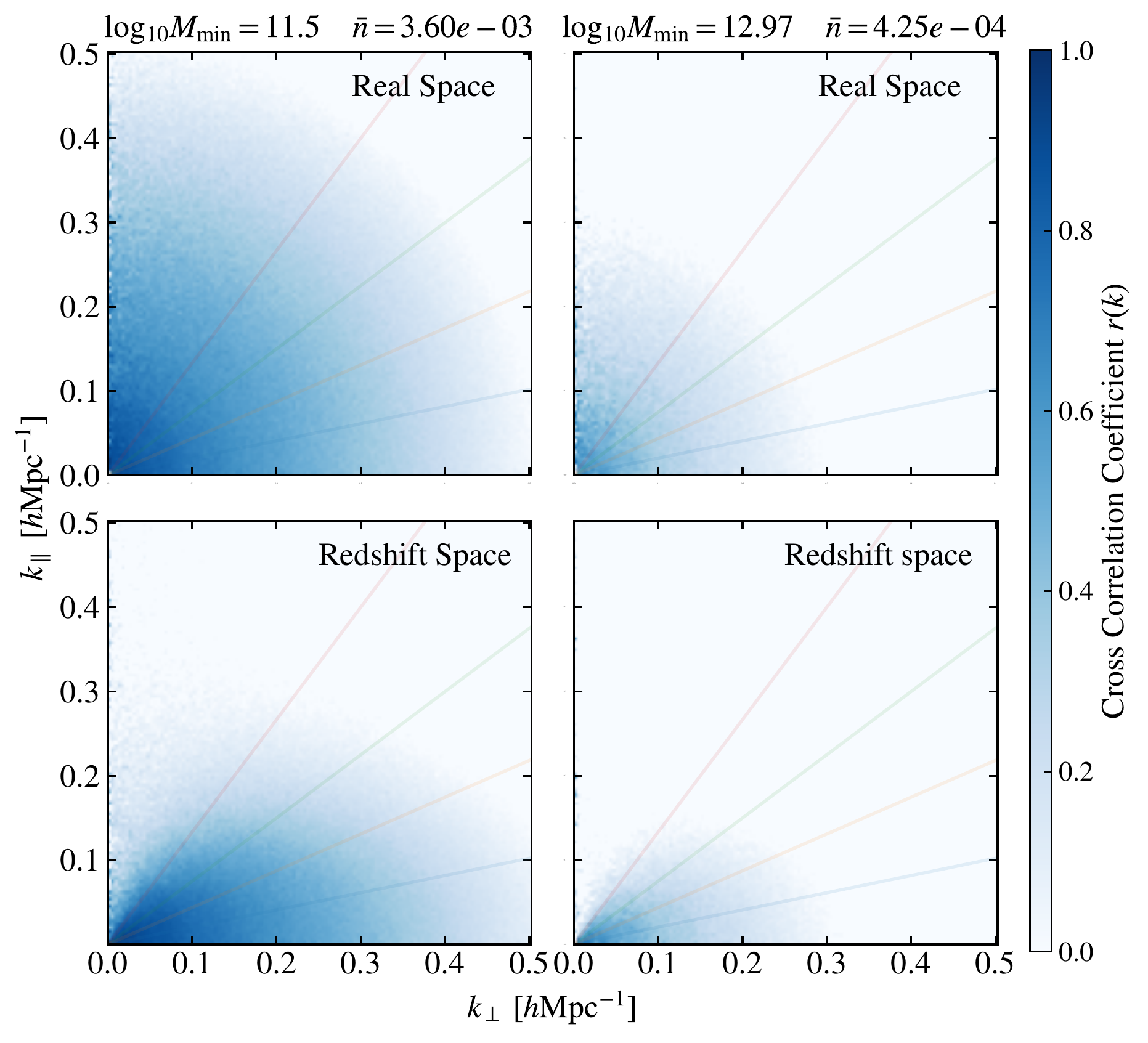}
    \caption{The two-dimensional correlation coefficient $r(k_{\perp}, k_{\parallel})$ of the full shear tidal reconstructed density field with the real space dark matter density field for two galaxy number densities $\bar{n}=3.6\times10^{-3}\ h^3\mathrm{Mpc}^{-3}$ and $4.25\times10^{-4}\ h^3\mathrm{Mpc}^{-3}$.
    The upper panels show real space results, while lower panels show results in redshift space.
    The light straight lines indicate five $\mu$ bins, from $0-0.2$, to $0.8-1.0$.
    In redshift space, the correlation coefficient is anisotropic, which becomes much smaller for modes with higher $\mu$ values.
    }
    \label{fig:Cor2D_5T}
\end{figure}

In Figure~\ref{fig:Cor2D_5T}, we plot the two-dimensional cross-correlation coefficients $r(k_{\perp}, k_{\parallel})$ between the full-shear reconstructed fields and the real space dark matter density field for two galaxy mock catalogs with $\bar{n}=3.6\times10^{-3}\ h^3\mathrm{Mpc}^{-3}$ and $4.25\times10^{-4}\ h^3\mathrm{Mpc}^{-3}$, respectively.
In general, the tidal reconstruction works better with the higher number density sample, i.e., lower shot noise, in both real and redshift spaces, because with a lower shot noise, the small-scale modes which dominate the reconstruction performance, are measured with a higher signal-to-noise ratio.
In real space, the correlation coefficients are isotropic for both higher and lower mass galaxy catalogs as expected.
For redshift space, we see that the correlation coefficient shows a clear angular dependence on the angle between the wave vector and the line of sight.
For reconstructed modes near the radial direction, i.e., $\mu\simeq1$, the correlation coefficient drops much faster than the transverse modes with $\mu\simeq0$ as the wave number increases.
Therefore, the reconstruction noise is much higher along the line-of-sight direction than the transverse plane, which is consistent with conclusions from the visual comparison in Figures~\ref{fig:Slice_5T_XY} and \ref{fig:Slice_5T_XZ}.

\begin{figure}[ht!]
    \centering
    \includegraphics[width=0.8\columnwidth]{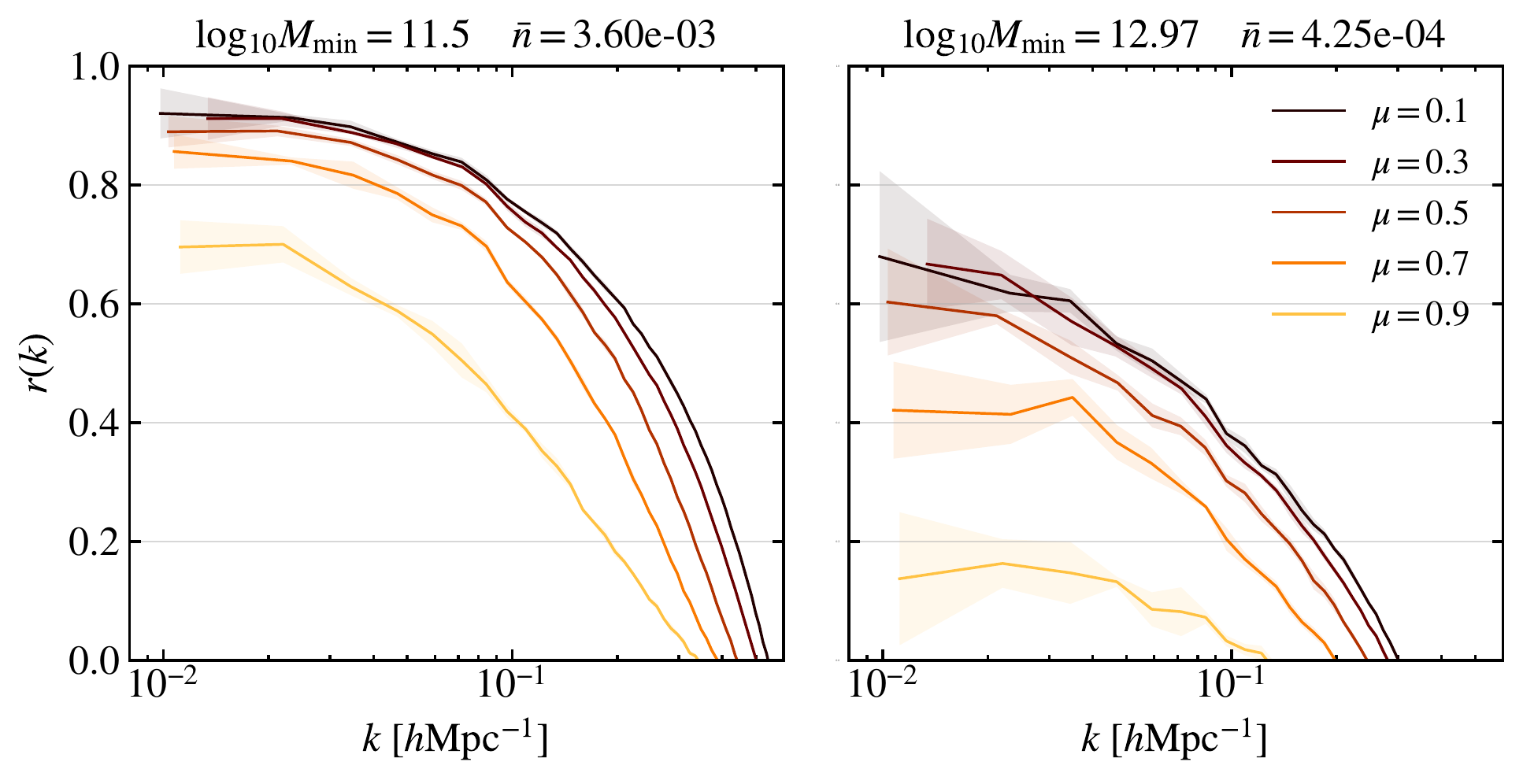}
    \caption{The cross-correlation coefficient $r(k,\mu)$ of the full-shear reconstructed fields with the dark matter density field, measured in five $\mu$ bins, $0-0.2$, $0.2-0.4$, etc, for two galaxy samples with $\bar{n}=3.6\times10^{-3}\ h^3\mathrm{Mpc}^{-3}$ and $4.25\times10^{-4}\ h^3\mathrm{Mpc}^{-3}$.
    While the reconstruction of modes in the highest $\mu$ bin is degraded by the RSD effect substantially, the other modes are reconstructed with high fidelity. 
    The envelops are the scatter estimated from the six independent realizations from the simulations.
    }
    \label{fig:R2DBin_RSD_5T}
\end{figure}

To see the anisotropy caused by the RSD effect more clearly, we plot the cross-correlation coefficients $r(k,\mu)$ measured in $(k,\mu)$ bins in Figure~\ref{fig:R2DBin_RSD_5T}.
The lines from dark to light show the correlation for five $\mu$ bins from 0.1 to 0.9.
The envelops are the scatter estimated from the six independent realizations from the simulations.
For the higher mass sample, i.e., $\bar{n}=4.25\times10^{-4}\ h^3\mathrm{Mpc}^{-3}$, the measured correlation coefficients have more fluctuations on large scales. 
This is not surprising as the reconstruction noise is much higher for this sample and the simulation boxes have limited volume and thus limited number of modes at scales $k\sim0.01\ h\mathrm{Mpc}^{-1}$.
The correlation coefficient can only reach $\sim0.7$, i.e., $1-r^2=P_N/P_{\delta_r\delta_r}\simeq0.5$, on the largest scales for the first few $\mu$ bins, $\mu=0.1$ and $0.3$.
Therefore, for the BOSS CMASS number densities $\bar{n}\sim4.25\times10^{-3}\ h^3\mathrm{Mpc}^{-3}$, the large-scale dark matter distribution can not be reconstructed with a high signal-to-noise ratio.
For the lower mass sample with a significantly higher number density of DESI galaxies, the correlation is higher than $\sim0.8$ at $k < 0.05\ h\mathrm{Mpc}^{-1}$, except the radial modes near the line of sight with $\mu=0.9$.
This allows the mapping of dark matter on large scales with a high signal-to-noise, which is highly beneficial for the multi-tracer analysis \citep[][]{2009JCAP...10..007M,2009MT}.

\begin{figure*}[ht!]
    \centering
    \includegraphics[width=0.8\columnwidth]{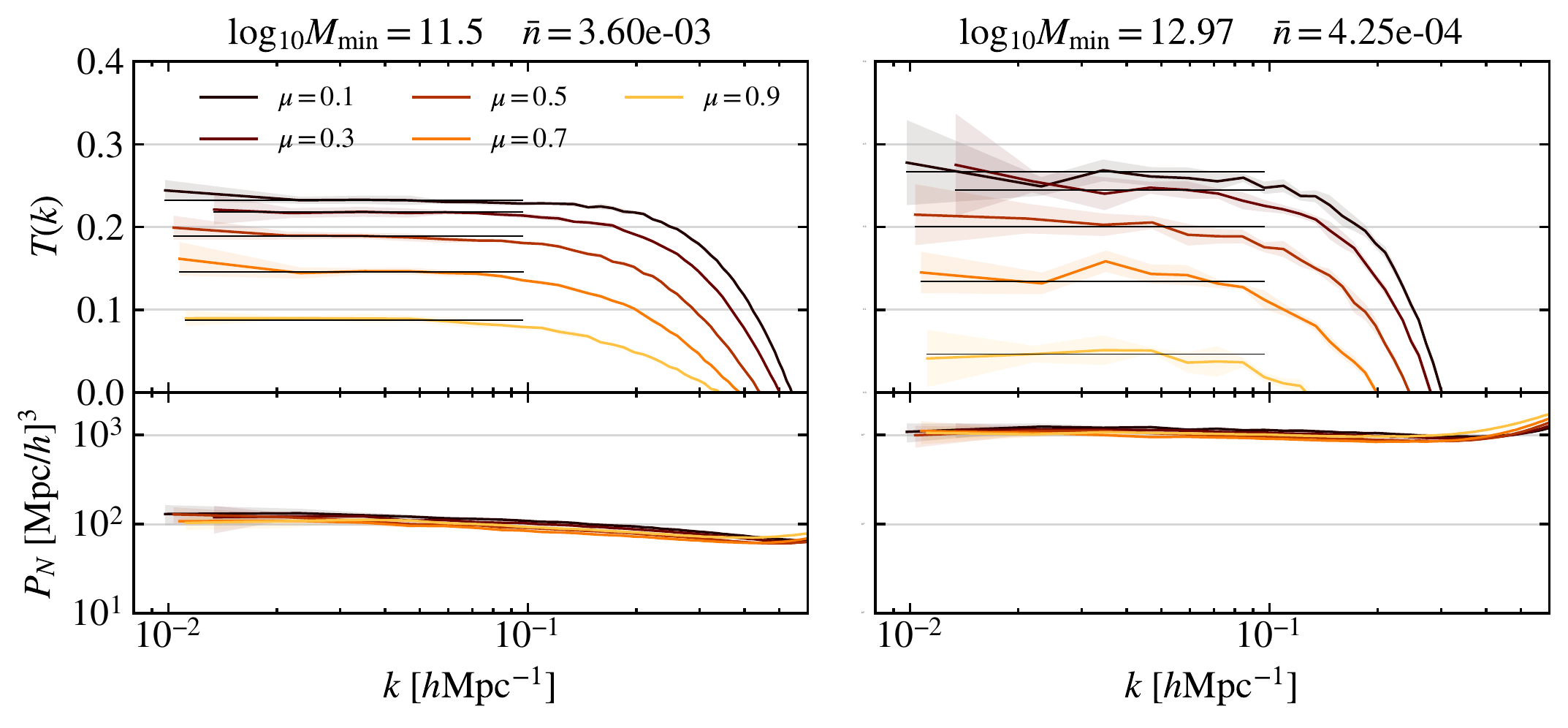}
    \caption{The propagator and reconstruction noise power spectrum for the full-shear reconstructed field in redshift space. 
    The horizontal solid lines show the simple model for the propagator at large scales.
    }
    \label{fig:TN2DBin_RSD_NT_5T}
\end{figure*}

Figure~\ref{fig:TN2DBin_RSD_NT_5T} 
presents the propagator and reconstruction noise power spectrum for tidal reconstruction in redshift space for two galaxy samples, measured in five $\mu$ bins from 0.1 to 0.9. 
The light shaded regions denote one standard deviation from repeating this estimate for the six simulations.
The propagator is defined as $T(k,\mu)=P_{\delta_r\delta}(k,\mu)/P_{\delta\delta}(k,\mu)$. 
Notice that the bias of the reconstructed field measured in this way is a function of $k$ and $\mu$. 
To disentangle the impact of RSD effect on tidal reconstruction, we apply the real space tidal shear estimator to redshift space mock galaxy density fields.
Therefore, the angular dependence of the propagator and noise power arises from the RSD effect alone.
From Figure~\ref{fig:TN2DBin_RSD_NT_5T}, for the lower mass galaxies, we notice that the propagator depends on the cosine $\mu$ with respect to the line of sight direction and its amplitude decreases as $\mu$ becomes larger.
We see a similar trend from the higher mass galaxy catalog though the measurements are noisier as the limited simulation volume and number of simulations.
The impact of the RSD effect manifests as different multiplicative biases on the amplitude of reconstructed modes with different cosine angles $\mu$ with respect to the radial direction, while in real space the propagator is isotropic without any dependence on the direction \citep[e.g.][]{2022ApJ...929....5Z}.
We will study this angular dependence in detail next.

The propagator approaches a constant value on large scales for all $\mu$ bins and deviates from the constant value at $k\gtrsim0.1\ h\mathrm{Mpc}^{-1}$.
It is more clear for the number density $\bar{n}=3.6\times10^{-3}\ h^3\mathrm{Mpc}^{-3}$, while less obvious for the lower number density $\bar{n}=4.25\times10^{-4}\ h^3\mathrm{Mpc}^{-3}$ due to the much higher reconstruction noise.
The similar trend has been observed in real space tidal reconstruction by \citet{2022ApJ...929....5Z}.
This is because the tidal shear estimators are derived in the squeezed limit, where the wavelength of the large-scale mode is much larger than that of the small-scale modes used for tidal reconstruction.
At the scale where the squeezed limit is assumed to break down, we would expect the bias for reconstructed fields begins running strongly with scale, as we see in Figure~\ref{fig:TN2DBin_RSD_NT_5T} \citep[see][for more discussions about this effect]{2022ApJ...929....5Z}.
This is the same as the CMB and 21~cm lensing estimators derived in the long wavelength limit \citep[see e.g.][for more discussions]{2008MNRAS.388.1819L,2010PhRvD..81l3015L,2012PhRvD..85d3016B,2019PhRvL.122r1301S}.

In Figure~\ref{fig:TN2DBin_RSD_NT_5T}, we also show the reconstruction noise power spectra for two galaxy samples.
We see that the noise power flattens on large scales.
This is as expected since in the squeezed limit, the reconstruction of large-scale modes is in the white homogeneous noise regime.
The reconstruction noise is much higher for the lower number density sample, since the higher shot noise enhances the stochasticity of reconstruction.
When the wavenumber is larger than $0.05\ h\mathrm{Mpc}^{-1}$, the noise power spectrum shows a mild disagreement with the white noise prediction on large scales.
We notice that the reconstruction noise power spectrum is more isotropic than the propagator. 
In the low-$k$ limit, the amplitude of the noise power spectrum can differ by only tens of percent for both simulated galaxy samples.

To use the tidal reconstructed field for cosmological inference, it is necessary to have an accurate description of the propagator and noise power spectrum.
The scale-dependent propagator or reconstruction bias is flat on large scales, with an angular dependence on the cosine with the line of sight.
The value of the propagator in the low-$k$ limit becomes smaller when the cosine $\mu$ becomes larger, being opposite to the linear Kaiser effect, where the power spectrum amplitude is larger near the line of sight, i.e., larger $\mu$ bins.
Therefore, we attempt to fit the large-scale propagator using a simple parametric form, 
\begin{equation}
    \label{eq:fit}
    T(k, \mu) = \beta_0 - \beta_2 \mu^2,
\end{equation}
to capture the angular dependent effect.
We then fit the propagator by minimizing the sum of squares 
\begin{equation}
    S=\sum_{i,j}\left(\hat{T}(k_i, \mu_j) - T(k_i, \mu_j)\right)^2,
\end{equation}
where the hat denotes measured data points from simulations.
Here we use the data points to $k_{\mathrm{max}} = 0.1\ h\mathrm{Mpc}^{-1}$ for all $\mu$ bins.
Note that the weight is uniform for all $k$ bins, while the power spectrum error usually scales as the inverse of number of modes in that $k$ bin.
This should be regarded as a particular weight to up-weight the estimated propagator on large scales, which avoids over-fitting at small scales that degrades the fit on large scales, i.e., the low-$k$ limit where the propagator approaches constant.
We use the \textsc{Scipy} routine {\tt scipy.optimize.curve\_fit} to implement the least square algorithm.
For the galaxy catalog with number density $\bar{n} = 3.6\times 10^{-3}\ h^3\mathrm{Mpc}^{-3}$, we obtain $\beta_0 = 0.234$ and $\beta_2 = 0.181$, with corresponding uncertainty $\sigma_{\beta_0} = 0.0012$ and $\sigma_{\beta_2} = 0.0027$.
For the higher mass sample with $\bar{n} = 4.25\times 10^{-4}\ h^3\mathrm{Mpc}^{-3}$, we obtain $\beta_0 = 0.269$ and $\beta_2 = 0.275$, with the variance $\sigma_{\beta_0} = 0.0028$, and $\sigma_{\beta_2} = 0.0063$, respectively.

We have plotted the best fit model for both high and low mass samples in Figure~\ref{fig:TN2DBin_RSD_NT_5T}.
For the higher number density sample, Equation~(\ref{eq:fit}) provides a fairly good description for the propagator at large scales.
While the measurement of the propagator is noisier for the lower number density sample, it can still be modeled by the simple two-parameter model within the one-sigma uncertainties on large scales.
The distinct angular dependence compared with the usual linear RSD effect, $b+f\mu^2$, is mainly due to that tidal reconstruction exploits small-scale structures, which is mostly impacted by the nonlinearities due to small-scale velocities, i.e., fingers of God effects.
One way to argue how well this linear bias model for tidal reconstruction works is to ask up to which scales $T(k,\mu)$ is a constant.
A significant scale dependence is a sign that the squeezed limit does not apply and higher order corrections must be included.
If we need to include higher $k$ modes in the cosmological analysis, the scale and angular dependence of $T(k,\mu)$ has to be modeled with high fidelity mock catalogs, which resembles the clustering properties of specific galaxy samples.

The reconstruction noise power spectrum is instead much more isotropic, with at most tens of percent fluctuations for different directions.
Since the ratio of noise power to total power is given by $1-r^2$, where $r$ is the cross-correlation coefficient, the reconstructed mode has been measured to the cosmic-variance dominated limit when the correlation coefficient is close to unity, $r\sim1$.
For high fidelity reconstruction, i.e., the higher number density sample, the noise power is subdominant on large scales, where $r>0.8$, except for the largest $\mu$ bin.
A ten percent variation in the noise power spectrum contributes to only a few percent of the total power spectrum.

Notice that the reconstruction noise power can not be directly compared with the shot noise prediction $1/\bar{n}$ for galaxies before reconstruction, since the propagator of reconstructed fields is generally $\sim0.2$, while the linear galaxy bias is of order $1\sim2$. 
The tidal reconstruction noise power spectrum is about $100\ h^{-3}\mathrm{Mpc}^3$ for the galaxy sample with $\bar{n} = 3.6\times 10^{-3}\ h^3\mathrm{Mpc}^{-3}$, and $10^3\ h^{-3}\mathrm{Mpc}^3$ for galaxies with $\bar{n} = 4.25\times 10^{-4}\ h^3\mathrm{Mpc}^{-3}$.
We could use the noise power divided by the propagator square, $P_N/T^2$ as a typical noise level for tidal reconstruction, i.e., about $100/0.2^2\ h^{-3}\mathrm{Mpc}^3=2.5\times10^3\ h^{-3}\mathrm{Mpc}^3$ for the low mass catalog and $10^3/0.2^2\ h^{-3}\mathrm{Mpc}^3=2.5\times10^4\ h^{-3}\mathrm{Mpc}^3$ for the high mass sample.
In general, the noise level is a few times larger than the shot noise of the halo catalogs used for reconstruction.
However, tidal reconstruction provides an independent tracer of the large-scale density, which can be used to cancel cosmic variance in the galaxy density.

In summary, the above results show tidal reconstruction method is very powerful at improving cosmological constraints using the sample variance cancellation technique \citep{2009MT,2009JCAP...10..007M}.
The simple parametric form of the propagator and nearly white isotropic reconstruction noise make the reconstructed tides field an ideal tracer for the multi-tracer method, especially for constraining the primordial non-Gaussianity \citep{2009MT,2021PhRvD.104l3520D}. 


\subsection{Transverse shear reconstruction}



Having discussed the full shear reconstruction, we now continue to explore the transverse shear reconstruction in redshift space. 
We have presented the density slices in the $x-y$ plane for transverse shear reconstruction in Figure~\ref{fig:Slice_5T_XY}.
The RSD changes little the reconstruction results, which is as expected since RSD does not affect transverse modes much.
Figure~\ref{fig:Slice_5T_XZ} shows the transverse shear reconstruction in the $x-z$ plane.
We note that the performance is nearly the same in both real and redshift spaces even for the radial modes.
This is due to that the RSD only affects the transverse shear indirectly.
This original proposal to avoid RSD effect, i.e., using only transverse shear $\gamma_1$ and $\gamma_2$ for tidal reconstruction, does work.

\begin{figure}[ht!]
    \centering
    \includegraphics[width=0.7\columnwidth]{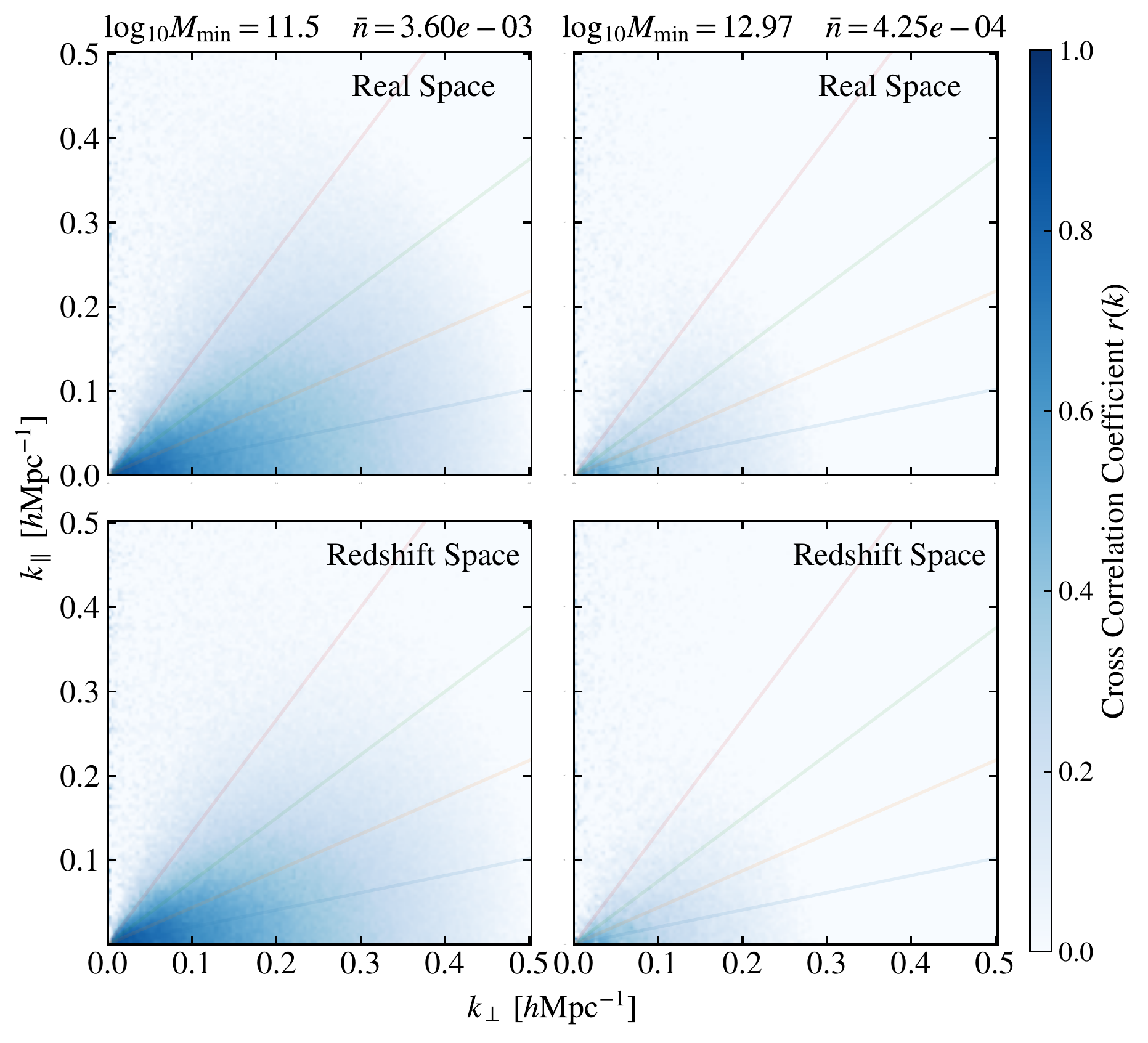}
    \caption{Same as Figure~\ref{fig:Cor2D_5T}, but for the transverse shear reconstruction method.
    The reconstruction shows a similar trend for both real and redshift spaces.}
    \label{fig:Cor2D_2T}
\end{figure}

\begin{figure}[ht!]
    \centering
    \includegraphics[width=0.8\columnwidth]{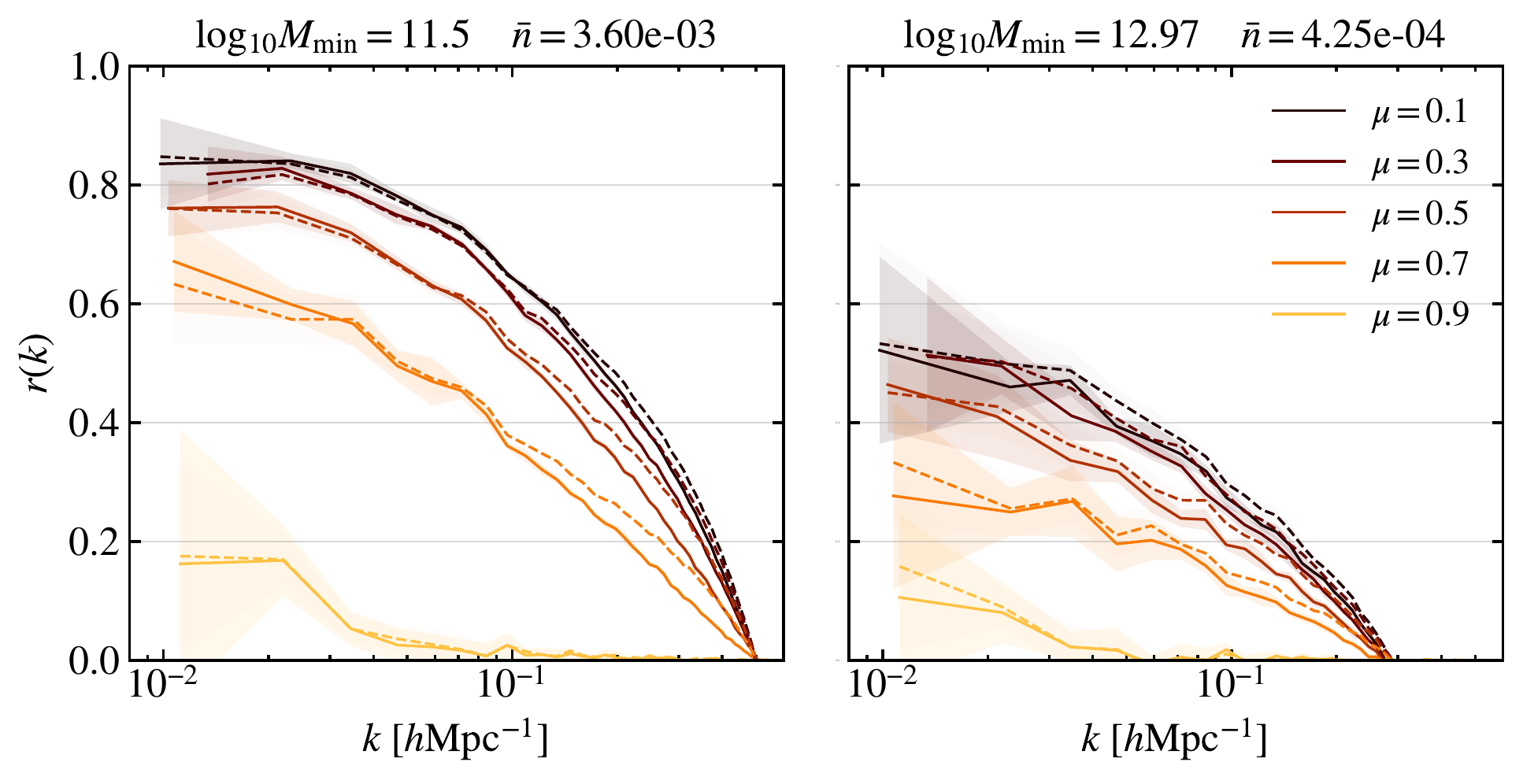}
    \caption{
    The cross-correlation coefficient for the transverse-shear reconstruction with two galaxy catalogs in redshift space (solid lines) and real space (dashed lines).
    }
    \label{fig:R2DBin_RSD_2T}
\end{figure}

Figure~\ref{fig:Cor2D_2T} shows the two-dimensional correlation coefficient between the transverse shear reconstructed fields and the original real space dark matter density field, for two galaxy number densities.
The correlation is much smaller for low $k_\perp$ and high $k_\parallel$ regime since these modes are inferred indirectly from the variation of the transverse shear $\gamma_1$ and $\gamma_2$ along the line of sight direction \citep{2012arXiv1202.5804P,2016PhRvD..93j3504Z,2019MNRAS.486.3864K}.
For both number densities, the correlation does not change much when the RSD effect is included.
The reconstruction shows a similar trend for both real and redshift spaces.
In Figure~\ref{fig:R2DBin_RSD_2T}, we plot the cross correlation coefficient measured in ($k,\mu$) bins.
The solid lines present the redshift space results while the dashed lines show the real space results.
For clarity, we only plot the $1\sigma$ error for the solid lines, but the errors are similar for both cases.
The correlation coefficient is almost the same for both real and redshift spaces, with some small discrepancies at small scales.
Therefore, for tidal reconstruction with only transverse shear $\gamma_1$ and $\gamma_2$, the mapping from real to redshift space is a second order effect.
This is consistent with the previous redshift space tidal reconstruction studied by
\citet{2019MNRAS.486.3864K}.

\begin{figure}[ht!]
    \centering
    \includegraphics[width=0.8\columnwidth]{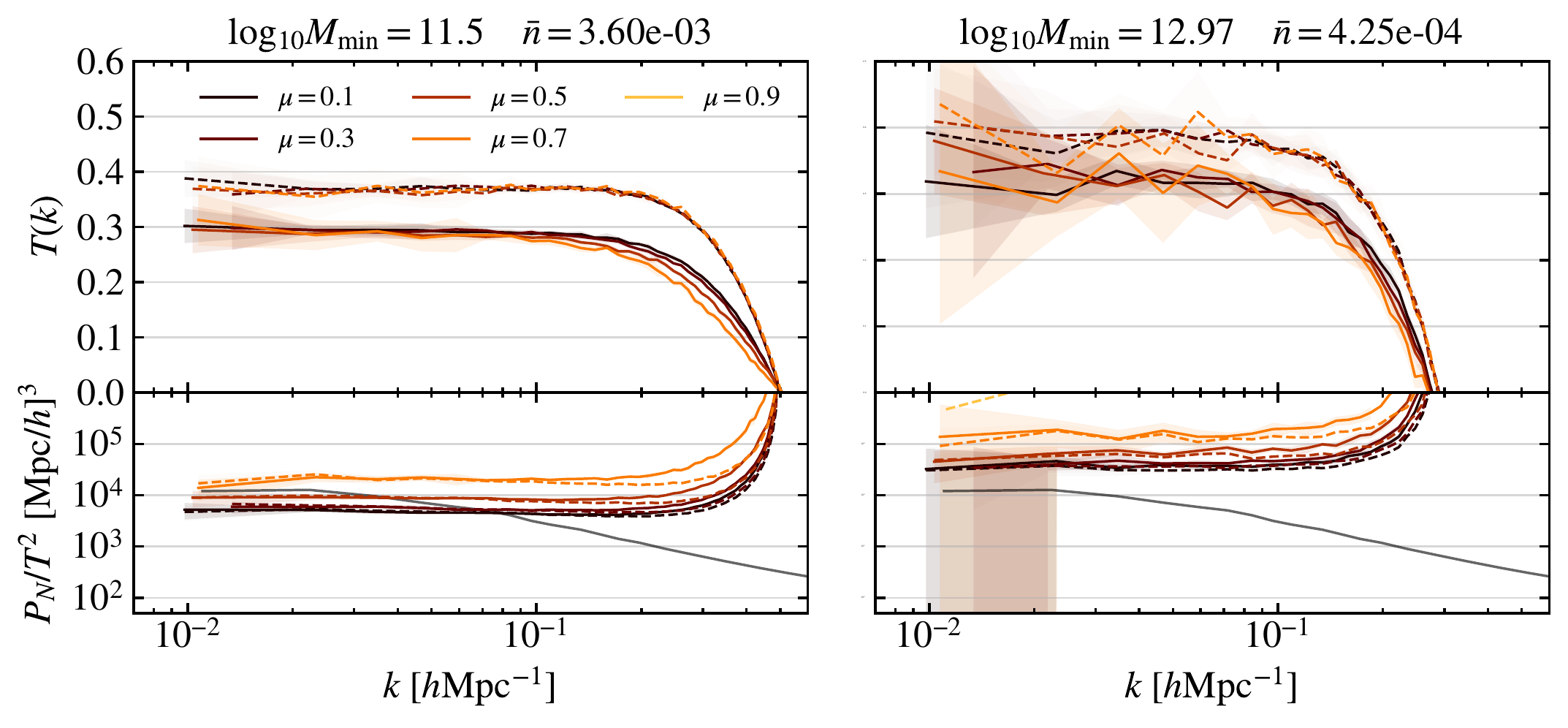}
    \caption{The propagator and reconstruction noise power spectrum for the transverse-shear reconstruction with two galaxy catalogs in redshift space (solid lines) and real space (dashed lines).
    Notice that we have plotted the normalized noise power spectrum $P_N/T^2$ to have a clear comparison between real and redshift space noises.
    The light grey lines in the bottom panels show the dark matter power spectrum for a better comparison with the noise power amplitude.
    }
    \label{fig:TN2DBin_RSD_2T}
\end{figure}

In Figure~\ref{fig:TN2DBin_RSD_2T}, we present the propagator and noise power spectrum of transverse shear reconstruction for two galaxy number densities.
The redshift space results are represented by solid lines while the real space results are plotted in dashed lines.
In real space, the propagator is isotropic and approaches a constant at large scales, even we only use the two transverse shear components for tidal reconstruction.
The salient feature is that the propagator is still nearly isotropic and scale-independent at large scales, $k<0.1\ h\mathrm{Mpc}^{-1}$, even in the presence of RSDs.
At small scales, $k>0.1\ h\mathrm{Mpc}^{-1}$, the RSD effect leads to a small angular-dependent feature, but not as apparent as the full-shear reconstruction algorithm.
Therefore, the RSD effect only changes the overall normalization of the reconstructed field, with a little anisotropic effect. 
In the low-$k$ limit, the propagator changes from $\sim0.4$ to $\sim0.3$ for the low mass sample and from $\sim0.5$ to $\sim0.4$ for the high mass sample.

To have a better comparison between the reconstruction noise, we have plotted the ratio of the noise power spectrum to the propagator, $P_N/T^2$ in Figure~\ref{fig:TN2DBin_RSD_2T}, since the absolute amplitude of noise power spectrum depends on the normalization of the reconstructed field as we discussed above.
This effectively corrects the normalization of the noise power spectrum on large scales, while increasing the small-scale noise power as the propagator becomes much smaller for higher wavenumber.
We omit the $\mu=0.9$ curve for $\bar{n}=4.25\times10^{-4}\ h^3\mathrm{Mpc}^{-3}$ which is greater than the upper limit of the plot. 
We see that the dashed and solid lines are nearly the same at large scales for both number densities, i.e., the reconstruction noise is almost the same with RSDs or not, as long as we normalize the noise power using the propagator.
Since we have $1-r^2=P_N/P_{\delta_r\delta_r}=P_N/T^2/(P_{\delta\delta}+P_N/T^2)$, i.e., an equal correlation coefficients $r$ leads to an equal noise power spectrum $P_N/T^2$, this conclusion is not surprising and directly follows from the result that cross correlation is the same in both cases as we have seen in Figure~\ref{fig:R2DBin_RSD_2T}.
We have plotted the real space dark matter power spectrum $P_{\delta\delta}$ for a direct comparison between $P_N/T^2$ and $P_{\delta\delta}$.
Since the propagator is isotropic in the low-$k$ limit, the RSD effect does not introduce additional angular dependence to the noise, except an overall scaling of the amplitude.
Thus, the reconstruction noise has a similar angular dependence as the noise in real space. 

While the transverse shear reconstruction has a higher and anisotropic noise compared to the full shear method, it has the advantage of being less impacted by errors in the galaxy redshift.
On large scales, the RSD mostly changes the overall normalization of the reconstructed field.
The propagator is still nearly isotropic in the presence of RSDs.
The anisotropy of reconstruction noise is largely due to the transverse nature of using only $\gamma_1$ and $\gamma_2$.
This demonstrates that the transverse shear reconstruction can be powerful for some cosmological applications which need to minimize the effect of redshift errors.
The modeling of reconstructed power spectrum could be accomplished using the real space results with a nuisance parameter describing the amplitude of the power spectrum.


\subsection{Exploration with the RSD effect}

\begin{figure}[ht!]
    \centering
    \includegraphics[width=0.7\columnwidth]{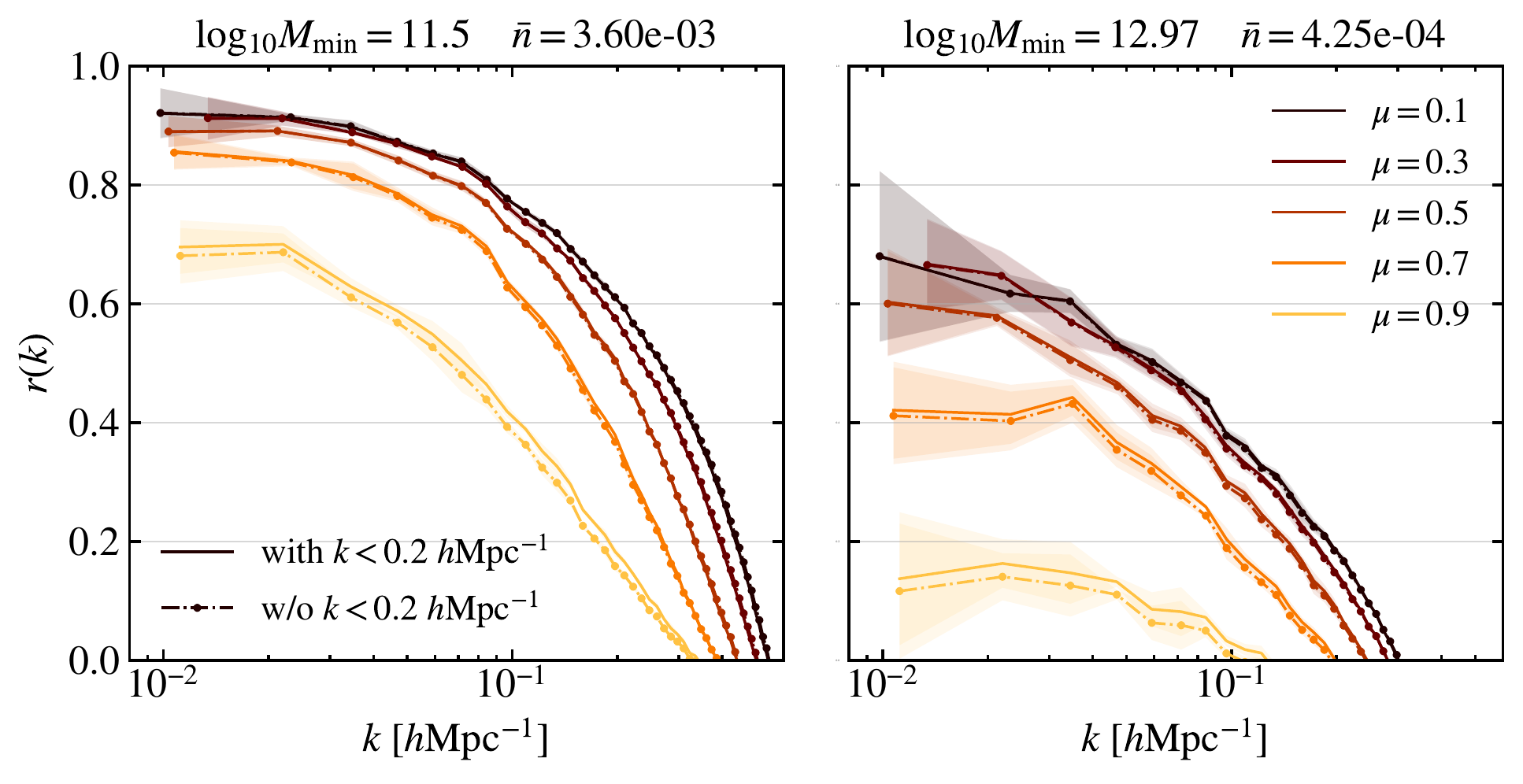}
    \caption{The cross-correlation coefficient for the full-shear reconstruction in the redshift space (solid lines) and the results without $k < 0.2 \ h\mathrm{Mpc}^{-1}$ modes (dash-dotted lines).}
    \label{fig:R2DBin_Cover_5T}
\end{figure}

There are two regimes in which we have a well-understanding of redshift-space distortions.
In the linear scales, a large-scale overdense region, towards which surrounding galaxies are falling, appears squashed in redshift space, which is known as the linear Kaiser effect \citep{1987MNRAS.227....1K}.
In Fourier space, the galaxy clustering is enhanced in redshift space than in real space by a factor $(b+f\mu^2)$. 
However, the linear distortions is only valid on large scales.
To evaluate the effect of linear distortions on tidal reconstruction, we can apply a high-pass filter to the galaxy overdensity used for reconstruction, which removes the large-scale modes where linear theory applies.
Figure~\ref{fig:R2DBin_Cover_5T} shows the cross-correlation coefficient between the reconstructed field and real space dark matter density for reconstruction without $k<0.2\ h\mathrm{Mpc}^{-1}$ modes.
We see that excluding all $k<0.2\ h\mathrm{Mpc}^{-1}$ modes only degrades the result a little.
If we exclude only $k<0.1\ h\mathrm{Mpc}^{-1}$ modes, one can hardly discern the difference between two curves.
We have confirmed that the propagator and noise power also change only a little when all $k<0.2\ h\mathrm{Mpc}^{-1}$ are excluded and almost no difference when excluding $k<0.1\ h\mathrm{Mpc}^{-1}$ modes.
This indicates that linear distortions have a negligible impact on tidal reconstruction, which makes sense since the reconstruction performance is dominated by the large number of small-scale modes.

This also explicitly demonstrates that the large-scale information from tidal reconstruction is independent of the original large-scale structures directly traced by galaxies, providing more information about cosmological parameters.
The modeling of the galaxy power spectrum in redshift space has advanced significantly and has been shown to be valid to $k\sim0.2-0.4\ h\mathrm{Mpc}^{-1}$, depending on specific methods \citep[see e.g.][]{2017JCAP...10..009H,2020PhRvD.102l3541N,2021JCAP...05..059S,2021JCAP...03..100C,2022MNRAS.514.3993P}, while most observable modes in galaxy surveys are still in the nonlinear regime and outside the realm of perturbative description.
Therefore, the tides information from nonlinear scales $k\sim1\ h\mathrm{Mpc}^{-1}$ is complementary with that from the large-scale power spectrum.
A multi-tracer analysis enables a potential improvement in the measurement of structure growth rate \citep{2009JCAP...10..007M}, similar to the $f_\mathrm{NL}$ constraints \citep{2021PhRvD.104l3520D}, which we plan to investigate in the future.

\begin{figure}[ht!]
    \centering
    \includegraphics[width=0.7\columnwidth]{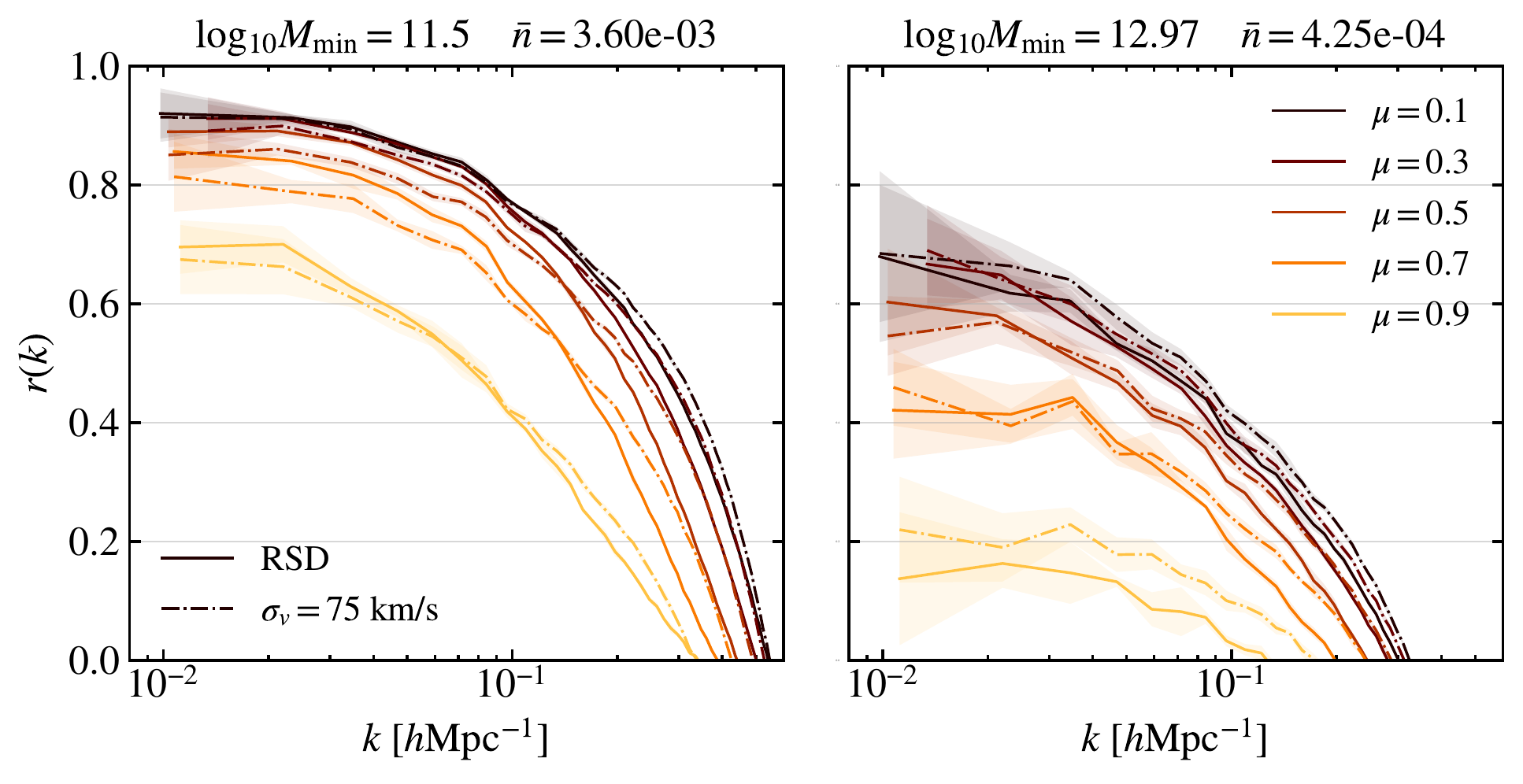}
    \caption{The cross-correlation coefficient for the full-shear reconstruction in redshift space (solid lines) and results for the synthetic redshift space galaxy catalogs which only includes the nonlinear velocity dispersion (dash-dotted lines).
    }
    \label{fig:R2DBin_PhotoZ_5T}
\end{figure}

As we move to smaller, nonlinear scales, the small-scale velocities elongate the galaxy clustering along the line of sight, usually known as fingers of God effect \citep{1972MNRAS.156P...1J}.
The quadrupole moment of the clustering at small scales has an opposite sign than it does in the linear case.
In Fourier space, the observed density field is damped in the radial direction.
This dominant small-scale nonlinearity in redshift space is caused by the nonlinear velocity dispersion $\sigma_v$ that has a different nature than the large-scale linear velocity.
To resemble the effect of nonlinear velocity dispersion, we move the galaxies along the line of sight with a random velocity drawn from a Gaussian distribution with standard deviation $\sigma_v$, instead of the real velocity of a galaxy from the simulation.
The importance of FoG is determined by the typical velocity dispersion $\sigma_v$, converted to comoving length units, $\sigma_\chi=(1+z)\sigma_v/H(z)$.
Figure~\ref{fig:R2DBin_PhotoZ_5T} shows the cross-correlation coefficient for tidal reconstruction with the synthetic redshift space galaxy catalogs which only include the small-scale random velocities.
We have tested a few values and found that with $\sigma_v = 75 \ \mathrm{km}/\mathrm{s}$, corresponding to $\sigma_\chi = 0.86 \ h^{-1}\mathrm{Mpc}$ at $z=0.6$, this small-scale velocity leads to a trend that qualitatively follows the real RSD effect. 
The cross-correlation coefficient also depends on the cosine with respect to the line of sight and its value also becomes smaller when we increase $\mu$.
We have confirmed that a larger $\sigma_v$ causes a larger degradation to tidal reconstruction, specifically a much smaller cross-correlation coefficient for high $\mu$ bins. 

\begin{figure}[ht!]
    \centering
    \includegraphics[width=0.8\columnwidth]{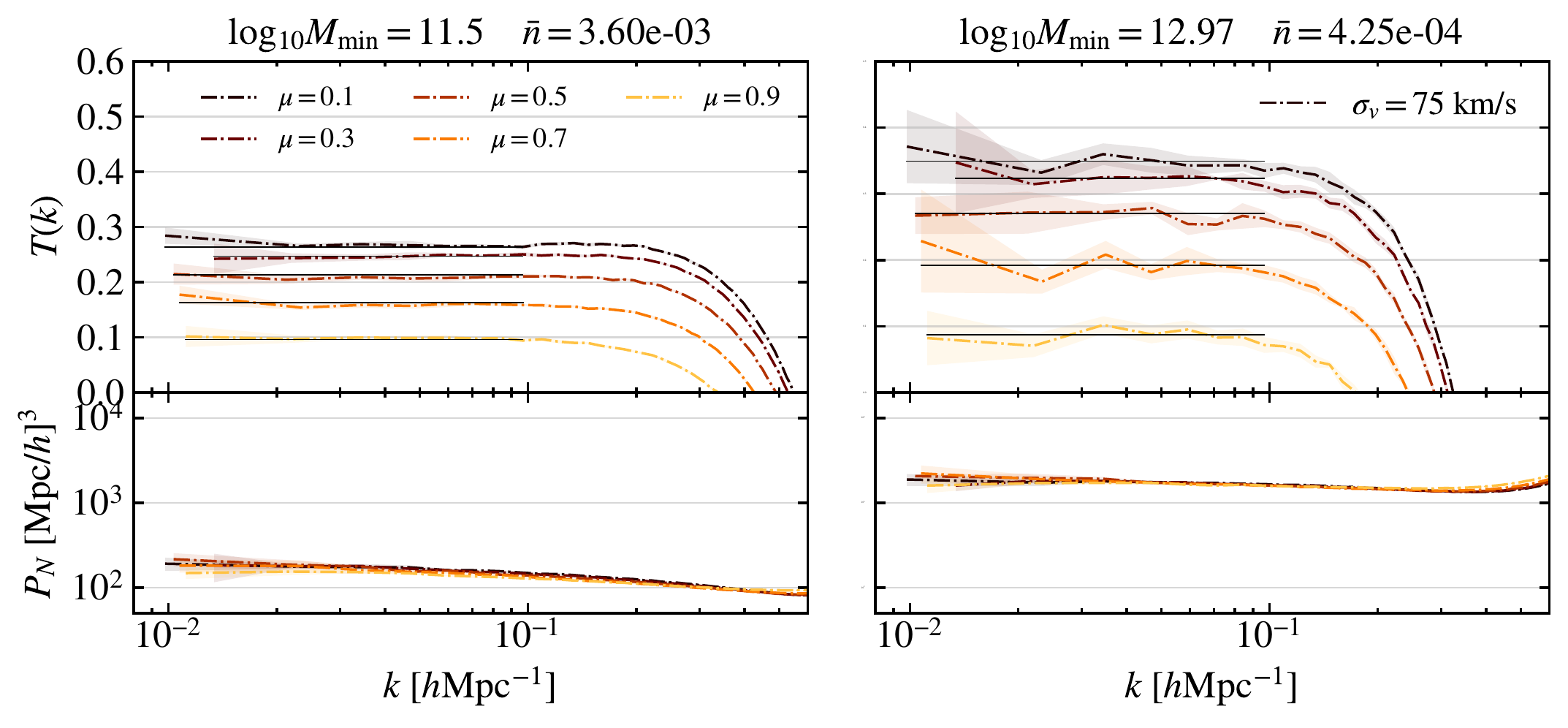}
    \caption{The propagator and reconstruction noise power spectrum for the full-shear tidal reconstruction with the synthetic redshift space galaxy catalogs which only include the nonlinear velocity dispersion.
    }
    \label{fig:TN2DBin_PhotoZ_NT_5T}
\end{figure}

In Figure~\ref{fig:TN2DBin_PhotoZ_NT_5T}, we present the propagator and noise power spectrum for tidal reconstruction with the synthetic catalogs which only include the small-scale velocities.
We find that the propagator can also be fitted using Equation~(\ref{eq:fit}), the simple two-parameter model $T(k, \mu) = \beta_0 - \beta_2 \mu^2$.
For the lower mass catalog with $\bar{n} = 3.6 \times 10^{-3}\  h^3\mathrm{Mpc}^{-3}$, we obtain $\beta_0 = 0.266$ and $\beta_2 = 0.210$, with the standard deviation $\sigma_{\beta_0} = 0.0015$ and $\sigma_{\beta_2} = 0.0035$.
For the higher mass galaxy sample with $\bar{n} = 4.25 \times 10^{-4}\  h^3\mathrm{Mpc}^{-3}$, we have $\beta_0 = 0.353$ and $\beta_2 = 0.328$, with the fitting error $\sigma_{\beta_0} = 0.0033 , \sigma_{\beta_2} = 0.0074$.
From this, it is clear that the dominant anisotropy of tidal reconstruction in redshift space is induced by the small-scale nonlinear velocity dispersion. 
The noise power spectrum is also nearly scale-independent in the low-$k$ limit where the long wavelength limit is a good approximation and has an angular fluctuation at tens of percent level between different $\mu$ bins.
This demonstrates that small-scale random nonlinear velocities can explain most behaviours we have observed for redshift space tidal reconstruction.

The value $\sigma_v = 75 \ \mathrm{km}/\mathrm{s}$ is compatible with the typical velocity dispersion of spectroscopic galaxy samples at the corresponding redshift \citep{2021JCAP...05..059S}.
While being much smaller than the large-scale linear bulk velocity \citep{2013PhRvD..87f3526Z,2013PhRvD..88j3510Z,2018PhRvD..97d3502Z,2018JCAP...09..006J}, the nonlinear velocity dispersion degrades the reconstruction of radial modes substantially and limits the information that we can extract from high values of $\mu$.
We have used the same $\sigma_v$ for both number densities.
This is simply due to that the reconstruction has a similar performance as the reconstruction with the real halo velocity estimated from the simulation.
In reality, the low mass sample should have a larger velocity dispersion due to additional satellite galaxies included.

The FoG effect produces a qualitatively correct result for the anisotropic reconstructed field, i.e., the angular dependence of the propagator and nearly isotropic noise.
However, we note that there are still discrepancies for cross-correlation coefficients and we cannot match the propagator and noise power spectrum between real and synthetic catalogs by adjusting only $\sigma_v$.
It is clear that higher order effects in the real to redshift space mapping need to be considered to have a full picture here.
However, accounting the full redshift-space distortions will be a tricky business.
We leave this for further work in the future.

\section{Discussion and Conclusion}
\label{sec:discussion} 

In this paper, we have applied the tidal reconstruction to redshift space galaxy fields from simulations, while most previous works focus on the real space reconstruction.
The large-scale density field can be recovered with high precision for the dense galaxy sample, with a correlation coefficient higher than 0.8 at the largest scales, $k<0.05\ h\mathrm{Mpc}^{-1}$, using the full shear method, except for the highest $\mu$ bin.
While for the sparse sample, the correlation coefficient can only reach $r\sim0.7$ at the large scales, limiting a substantial improvement in cosmological parameter measurement using the sample variance cancellation technique.
Although for the existing galaxy samples such as SDSS BOSS/eBOSS \citep[][]{2017MNRAS.470.2617A,2021PhRvD.103h3533A}, the number density is insufficient for tidal reconstruction to be efficient, i.e., close to the number density $\bar{n}=4.25\times10^{-3}\ h^3\mathrm{Mpc}^{-3}$ we have studied in this paper, the ongoing and future surveys will have a much higher number density such as DESI BGS and ELG \citep[][]{2022arXiv220808512H,2022arXiv220808513R}, Euclid \citep[][]{2018LRR....21....2A,2020A&A...642A.191E}, SPHEREx \citep[][]{2014arXiv1412.4872D}, MegaMapper \citep[][]{2019BAAS...51g.229S,2019BAAS...51c..72F}, etc (See \citealt{2021JCAP...12..049S,2022arXiv220307506F} for a review).
High density galaxy clustering and Stage-5 surveys are also being planned \citep[][]{2022arXiv220307291D,2022arXiv220903585S}.
The tidal reconstruction method allows a substantial improvement on the cosmological parameter constraints, e.g., local primordial non-Gaussianity, given the fixed survey volume and galaxy number density, at no additional cost.
This makes tidal reconstruction a promising probe of cosmology.

The small-scale FoG effect, i.e., nonlinear velocity dispersion, leads to a degradation to the full shear reconstruction in redshift space, especially for high $\mu$ values.
This makes sense since tidal reconstruction performance is dominated by the small-scale density modes.
However, as the transverse shear terms are only indirectly affected by the real to redshift space mapping, the transverse shear method is largely insensitive to the RSD.
Therefore, while being noisier than the full shear method, the transverse shear reconstruction could still be useful in some certain cases.

Tidal reconstruction acquires a large-scale linear bias, which is constant to an excellent approximation.
In redshift space, for full shear reconstruction this bias becomes angular dependent due to the anisotropic nature of redshift space galaxy density field.
However, we find that the reconstruction bias can be well described by a simple two-parameter model on large scales.
The noise power spectrum is nearly isotropic and scale-independent at $k<0.05\ h\mathrm{Mpc}^{-1}$.
Thus, we expect that for modes with $k<0.05\ h\mathrm{Mpc}^{-1}$, the noise power spectrum can be modeled as a constant term to a good approximation in the cosmological data analysis.
Therefore, this makes it possible to use the reconstructed modes alongside directly observed galaxy density modes to constrain $f_\mathrm{NL}$ using an effective multi-tracer approach \citep{2021PhRvD.104l3520D}.
However, in order to reach the theoretical threshold between single and multi-field inflation models $f_\mathrm{NL}\sim1$ \citep[see e.g.][]{2014arXiv1412.4671A}, more detailed studies using very large volume simulations with primordial non-Gaussianity will be needed since even percent level stochasticities can significantly impact the inference of $f_\mathrm{NL}$. 
We plan to study this in future.

The propagator shows a characteristic anisotropy in the cosine $\mu$, $T(k,\mu)=\beta_0-\beta_2\mu^2$, which mostly arises from the nonlinear velocity dispersion as we have shown above.
It might be possible to derive this characteristic scaling in $\mu$ analytically in the large-scale limit, by assuming a tidal coupling or response function in the long wavelength limit.
This response function could be obtained from tides simulations \citep[see e.g.][]{2018MNRAS.479..162S,2021MNRAS.503.1473S,2020MNRAS.496..483M,2021JCAP...04..041A,2021MNRAS.504.1694R}.
We plan to investigate this topic in a future work.
While being noisy for high $\mu$ values, tidal reconstruction can obtain a high signal-to-noise reconstruction of smaller $\mu$ modes.
This is highly complementary with other reconstruction methods such as the kinetic Sunyaev-Zel'dovich velocity reconstruction \citep[see e.g.][]{2018arXiv181013423S,2021arXiv211111526C,2022JCAP...09..028G}, where the radial modes with $\mu\sim1$ have the lowest noise.

One of the major applications of tidal reconstruction is to recover the lost radial modes due to foreground in 21~cm intensity mapping surveys.
The neutral hydrogen maps have much smaller fingers of God effects than the typical spectroscopic galaxy samples at the same redshift, driven by a small number of satellite galaxies with a smaller velocity dispersion \citep{2018ApJ...866..135V,2022arXiv220712398O}.
Therefore, it may be even more beneficial for 21~cm surveys such as CHIME \citep{2022ApJS..261...29C}, HIRAX \citep{2022JATIS...8a1019C}, stage-II experiments \citep{2018arXiv181009572C}, PUMA \citep{2019BAAS...51g..53S,2020arXiv200205072C}, etc.
However, further studies are required to study the effects of foreground contamination and instrumental effects which we leave to future work. 

The isotropic modulation to the local power spectrum can also give an estimate of the large-scale density by measuring the amplitude of the small-scale power spectrum in different subvolumes \citep[e.g.][]{2014JCAP...05..048C,2014PhRvD..90j3530L,2015JCAP...09..028C} or using a tidal field quadratic estimator as we presented here.
However, compared with the local anisotropic distortions, the isotropic modulation is more likely to be impacted by observational systematics, e.g., variations in the foreground stars, seeing, and galactic dust extinction, since both lead to the change in local galaxy power spectrum.
A detailed exploration of observational systematics will be presented in a future paper.

The reconstructed field is quadratic in the small-scale galaxy density.
The cross spectrum of the reconstructed field with the original galaxy field is a bispectrum of the galaxy density, while the power spectrum of the reconstructed field is a trispectrum.
Therefore, we are using higher order statistics, 3-point function and 4-point function, in redshift space to improve cosmological measurements.
There are other similar methods using quadratic functions of the density field to exploit higher-order information such as the skew power spectrum \citep[see e.g.][]{2015PhRvD..91d3530S,2020JCAP...04..011M,2020JCAP...08..007D,2021JCAP...03..020S}, but requires a perturbative description.
However, the perturbation theory has limited range of validity and eventually breaks down in the nonlinear regime $k\sim1\ h\mathrm{Mpc}^{-1}$.
For future large-scale structure studies,
we are sensitive to the breakdown of perturbation theory, therefore, it is of great importance to further develop methods that could efficiently exploit nonlinear information in redshift space beyond the linear theory \citep{2021JCAP...10..044F}.

\section*{Acknowledgement}

\begin{acknowledgments}

Ue-Li Pen receives support from Ontario Research Fund-Research Excellence Program (ORF-RE), Natural Sciences and Engineering Research Council of Canada (NSERC) [funding reference number RGPIN-2019-067, CRD 523638-18, 555585-20], Canadian Institute for Advanced Research (CIFAR), Canadian Foundation for Innovation (CFI), the National Science Foundation of China (Grants No. 11929301), Thoth Technology Inc, Alexander von Humboldt Foundation, and the National Science and Technology Council (NSTC) of Taiwan (111-2123-M-001-008-, and 111-2811-M-001-040-).
Computations were performed on the Niagara supercomputer at the SciNet HPC Consortium and the SOSCIP Consortium's CPU computing platform.
SciNet is funded by: the Canada Foundation for Innovation; the Government of Ontario; Ontario Research Fund - Research Excellence; and the University of Toronto \citep[][]{Loken_2010}.
SOSCIP is funded by the Federal Economic Development Agency of Southern Ontario, the Province of Ontario, IBM Canada Ltd., Ontario Centres of Excellence, Mitacs and 15 Ontario academic member institutions.
\end{acknowledgments}

\vspace{5mm}

\bibliography{sample631}{}
\bibliographystyle{aasjournal}

\end{document}